\documentclass[nofootinbib,10pt,twocolumn,preprintnumbers]{revtex4-1}
\usepackage{amsmath,amssymb,pgf,pgfarrows,pgfnodes,float,appendix, hyperref,slashed,breakurl,graphicx}
\definecolor{Color}{rgb}{0.28, 0.24, 0.55}
\definecolor{Orange}{rgb}{1,0.38,0.11}
\hypersetup{
    colorlinks = true,
    citecolor = Orange,
    linkcolor = Color,
    urlcolor  = Orange,
}
\definecolor{internationalorange}{rgb}{1.0, 0.31, 0.0}

\usepackage{graphicx}
\usepackage{subfigure}
\usepackage[margin=0.9in]{geometry}

\usepackage{enumerate}

\usepackage{amsmath}

\newcommand{\SU}{\text{SU}}

\usepackage{color, colortbl}
\definecolor{Gray}{gray}{0.8}
\definecolor{GrayLight}{gray}{0.4}
\definecolor{Darkgreen}{RGB}{30,120,30}
\definecolor{granate}{rgb}{0.8039,0.2,0.2}
\newcommand{\beq}{\begin{equation}}
\newcommand{\eeq}{\end{equation}}
\newcommand{\bea}{\begin{eqnarray}}
\newcommand{\eea}{\end{eqnarray}}

\usepackage{eso-pic}

\usepackage{xparse}

\usepackage{tikz}
\usetikzlibrary{"arrows", "automata", "backgrounds", "calendar", "chains", "matrix", "mindmap", "patterns", "petri", "shadows", "shapes.geometric", "shapes.misc", "spy", "trees"}
\usetikzlibrary{arrows,shapes}
\usetikzlibrary{trees}
\usetikzlibrary{matrix,arrows} 				
\usetikzlibrary{positioning}				
\usetikzlibrary{calc,through}				
\usetikzlibrary{decorations.pathreplacing}  
\usepackage{pgffor}							

\usetikzlibrary{decorations.pathmorphing}	
\usetikzlibrary{decorations.markings}
\tikzset{
    vector/.style={decorate, decoration={snake}, draw},
	provector/.style={decorate, decoration={snake,amplitude=2.5pt}, draw},
	antivector/.style={decorate, decoration={snake,amplitude=-2.5pt}, draw},
    fermion/.style={draw=black, postaction={decorate},
        decoration={markings,mark=at position .55 with {\arrow[draw=black]{>}}}},
    fermion1/.style={draw=black, postaction={decorate},
        decoration={markings,mark=at position .25 with {\arrow[draw=black]{>}}}},
    fermioncyan/.style={draw=black, postaction={decorate},
        decoration={markings,mark=at position .55 with {\arrow[draw=cyan]{<}}}},
        fermionp/.style={draw=black, postaction={decorate},
        decoration={markings,mark=at position .78 with {\arrow[draw=black]{>}}}},
    fermionpp/.style={draw=black, postaction={decorate},
        decoration={markings,mark=at position 0.27 with {\arrow[draw=black]{>}}}},
    fermionppp/.style={draw=black, postaction={decorate},
        decoration={markings,mark=at position 1.0 with {\arrow[draw=black]{>}}}},
    fermiondif/.style={draw=black, postaction={decorate},
        decoration={markings,mark=at position .7 with {\arrow[draw=black]{>}}}},
            fermiondif2/.style={draw=black, postaction={decorate},
        decoration={markings,mark=at position .7 with {\arrow[draw=black]{<}}}},
    fermionend/.style={draw=black, postaction={decorate},
        decoration={markings,mark=at position 1 with {\arrow[draw=black]{>}}}},
    fermionuchannel2/.style={draw=black, postaction={decorate},
        decoration={markings,mark=at position .4 with {\arrow[draw=black]{>}}}},
    scalardif/.style={dashed,draw=black, postaction={decorate},
        decoration={markings,mark=at position .7 with {\arrow[draw=black]{>}}}},
    scalarend/.style={dashed,draw=black, postaction={decorate},
        decoration={markings,mark=at position 1 with {\arrow[draw=black]{>}}}},
    fermionbar/.style={draw=black, postaction={decorate},
        decoration={markings,mark=at position .55 with {\arrow[draw=black]{<}}}},
    fermionnoarrow/.style={draw=black},
    gluon/.style={decorate, draw=black,
        decoration={coil,amplitude=4pt, segment length=5pt}},
    scalar/.style={dashed,draw=black, postaction={decorate},
        decoration={markings,mark=at position .55 with {\arrow[draw=black]{>}}}},
    scalarloop/.style={dashed,draw=black, postaction={decorate},
        decoration={markings,mark=at position .75 with {\arrow[draw=black]{<}}}},
            scalarloop1/.style={dashed,draw=black, postaction={decorate},
        decoration={markings,mark=at position .25 with {\arrow[draw=black]{<}}}},
    scalarcyan/.style={dashed,draw=black, postaction={decorate},
        decoration={markings,mark=at position .55 with {\arrow[draw=cyan]{>}}}},
    scalaruchannel1/.style={dashed,draw=black, postaction={decorate},
        decoration={markings,mark=at position .7 with {\arrow[draw=black]{>}}}},
                  scalaruchannel2/.style={dashed,draw=black, postaction={decorate},
        decoration={markings,mark=at position .4 with {\arrow[draw=black]{>}}}},
    scalarbar/.style={dashed,draw=black, postaction={decorate},
        decoration={markings,mark=at position .55 with {\arrow[draw=black]{<}}}},
    scalarnoarrow/.style={dashed,draw=black},
    electron/.style={draw=black, postaction={decorate},
        decoration={markings,mark=at position .55 with {\arrow[draw=black]{>}}}},
	bigvector/.style={decorate, decoration={snake,amplitude=4pt}, draw},
}

\NewDocumentCommand\semiloop{O{black}mmmO{}O{above}}
{%
\draw[#1] let \p1 = ($(#3)-(#2)$) in (#3) arc (#4:({#4+180}):({0.5*veclen(\x1,\y1)})node[midway, #6] {#5};)
}

\tikzstyle{block} = [draw, rectangle, 
    minimum height=3em, minimum width=6em]

\usepackage{tikz}
\usetikzlibrary{fit}
\tikzset{%
  highlight/.style={rectangle,rounded corners,color=granate,draw,text opacity =1,
    fill opacity=0.5,thick,inner sep=0pt}
}

\usepackage{placeins}

\tikzset{
    cross/.pic = {
    \draw[rotate = 45] (-#1,0) -- (#1,0);
    \draw[rotate = 45] (0,-#1) -- (0, #1);
    }
}

\tikzset{
    square/.style={%
        draw=none,
        circle,
        append after command={%
            \pgfextra \draw[#1] (\tikzlastnode.north-|\tikzlastnode.west) rectangle 
                (\tikzlastnode.south-|\tikzlastnode.east);\endpgfextra}
    },
    square/.default=black
}

\tikzstyle{block} = [draw, rectangle, 
    minimum height=3em, minimum width=6em]

\begin{document}

\preprint{CALT-TH/2023-004}

\title{\Large {\bf{Automatic Nelson-Barr solutions to the strong CP puzzle}}}
\author{Pavel Fileviez P\'erez$^{1}$, Clara Murgui$^{2}$, Mark B. Wise$^{2}$}
\affiliation{$^{1}$Department of Physics and Center for Education and Research in Cosmology and Astrophysics (CERCA), Case Western Reserve University, Cleveland, OH 44106, USA \\
$^{2}$Walter Burke Institute for Theoretical Physics, California Institute of Technology, Pasadena, CA 91125, USA}


\begin{abstract}
We discuss a simple model, based on the gauge group $\SU(3)_C\otimes \SU(2)_L \otimes \text{U}(1)_Y\otimes \text{U}(1)_R$, where the Nelson-Barr solution to the strong CP problem is implemented. This model automatically provides a high quality solution to  the strong CP puzzle. Weak CP violation in the lepton sector arises in the same fashion as in the quark sector.
We derive explicit expressions for the flavor changing couplings of the electroweak and Higgs bosons. These expressions are more general than the particular model considered. Constraints from finite naturalness are briefly discussed. We briefly also discuss  related models based on the gauge group B-L. 
\end{abstract}

\maketitle

\section{Introduction}
In the standard model (SM) 
the CP symmetry is explicitly broken by the QCD vacuum angle 
and by the interactions of the left-handed quarks with the $W$-gauge bosons.  
The QCD vacuum angle, $\theta_{\rm QCD}$, appears in the CP-violating term
\begin{equation}
{\cal L}_{\rm QCD} \supset  \frac{\alpha_s \theta_{\rm QCD}}{16 \pi} G_{\mu \nu}^a \epsilon^{\mu \nu \rho \sigma} G_{\rho \sigma}^{a} ,
\end{equation}
where $a=1..8$, $\alpha_s$ is the strong coupling and $G_{\mu \nu}^{a}$ is the field strength tensor for gluons.
The basis invariant parameter that enters in the neutron electric dipole moment is given by
\begin{equation}
\bar{\theta}_{\rm QCD} = \theta_{\rm QCD} + \text{arg} \{ \text{Det} ({\cal M}_u {\cal M}_d)\},
\label{eq:thetabar}
\end{equation}  
where the last term is the contribution from CP-violating phases in the quark mass matrices.
The vacuum angle $\bar{\theta}_{\rm QCD}$ must be very small, $\bar{\theta}_{\rm QCD}< 10^{-10}$~\cite{Abel:2020pzs}, in order to satisfy the 
experimental bounds on the neutron electric dipole moment. The second term in Eq.~\eqref{eq:thetabar} is related 
to the large CP-violating phase in the Cabibbo-Kobayaski-Maskawa (CKM) matrix and an awkward cancellation between it  and ${\theta}_{\rm QCD}$  is required for ${\bar \theta}_{\rm QCD}$ to be very small. This is the strong CP problem. 

There are two well studied solutions to the strong CP-problem:  
the Peccei-Quinn (PQ) mechanism~\cite{Peccei:1977hh,Peccei:1977ur} and the Nelson-Barr (NB) mechanism~\cite{Nelson:1983zb,Barr:1984qx,Nelson:1984hg}. In the case of the 
PQ mechanism one postulates the existence of an anomalous global ${\rm U}(1)_{\rm PQ}$ symmetry 
that is spontaneous broken and gives rise to the existence of the axion~\cite{Wilczek:1977pj,Weinberg:1977ma,Kim:1979if,Shifman:1979if,Dine:1981rt,Zhitnitsky:1980tq}, a possible dark matter candidate~\cite{Preskill:1982cy,Abbott:1982af,Dine:1982ah}. 
When the Peccei-Quinn symmetry is an automatic consequence of a spontaneously broken gauge symmetry it is difficult to get a small enough value of ${\bar \theta}_{\rm QCD}$ since operators of very high dimension must also preserve the symmetry  \cite{Georgi:1981pu}. The degree to which higher dimension operators preserve the smallness of ${\bar \theta}_{\rm QCD}$ is called the quality of the solution. 

In the case of the Nelson-Barr mechanism CP is spontaneously 
broken and new quarks are added in such way that the $\text{arg} \{ \text{Det} ({\cal M}_u {\cal M}_d)\}$ vanishes at tree level.\footnote{For a related mechanism that predates Nelson-Barr see Ref.~\cite{Mohapatra:1978fy}. An implementation of this idea is given in Ref.~\cite{Babu:1989rb}.} Nonetheless,  a large CP violating  CKM phase is permitted. See  Bento, Branco and Parada in Ref.~\cite{Bento:1991ez} for a discussion of the CKM matrix. The quality of the solution is easier to ensure in NB models because forbidding dimension five operators can be sufficient.

Gauge extensions of the SM  that implement the NB mechanism have been constructed.  
For previous studies see Refs.~\cite{Frampton:1990pr,Schwichtenberg:2018aqc,Valenti:2021xjp,Asadi:2022cco}. The spontaneous breaking of CP gives rise to domain walls which will dominate the energy density of the universe unless inflation~\cite{Guth:1980zm,Linde:1981mu,Albrecht:1982wi} (or whatever solves the horizon problem) occurs after the spontaneous breaking of CP~\cite{Vilenkin:1984ib,McNamara:2022lrw}.
As Ref.~\cite{Asadi:2022cco} pointed out, a new gauge symmetry can help increase the quality of the NB mechanism by forbidding the lowest higher dimensional CP violating operators. Assuming that CP invariance is restored at high temperatures this relaxes the upper bound on the CP breaking scale which otherwise would be in tension with the simplest models of inflation. Note that there are cases where CP is not restored at high temperatures~\cite{Dvali:1996zr}. 

In this article, we discuss simple gauge theories based on the gauge group ${\cal G}_{\rm SM} \otimes {\rm U}(1)_{R}$ (see for example, \cite{Appelquist:2002mw}), 
where ${\cal G}_{\rm SM}$ is the SM gauge group. In these models the NB mechanism is an automatic consequence of the gauge theory and matter content. The gauge symmetry increases the quality of the NB mechanism forbidding  dimension five operators that can spoil it. This relieves tension between solving the strong CP problem and cosmology. In this model, CP-violation in the lepton sector arises in the
same fashion as in the quark sector.
We also briefly discuss models based on ${\cal G}_{\rm SM} \otimes {\rm U}(1)_{B-L}$.

This article is organized as follows: In section II, we discuss the gauge theories for spontaneous CP-violation that we study. 
In section III, we discuss the implementation of the Nelson-Barr mechanism and also the quality of the solution to the strong CP problem.
In section IV, we review how to obtain a realistic CKM and PMNS matrices. 
Flavor violating interactions are discussed in section V, while the bounds from finite naturalness are discussed in section VI. We summarize our main findings in section VII.
%
\section{Theoretical Framework}
A gauge theory for spontaneous CP violation (SCPV) that implements automatically the Nelson-Barr mechanism can be constructed using the gauge symmetry group:
\begin{equation}
{\cal G}_{R} = \SU(3)_C \otimes \SU(2)_L \otimes  \text{U}(1)_Y \otimes \text{U}(1)_R.
\end{equation}
The SM fermion content (plus three right-handed neutrinos) with the right-handed fields carrying a charge $r$,
\begin{equation}
\begin{aligned}
q_L &\sim (3,2,1/6,0), & \hspace{-0.2 cm} \ell_L &\sim (1,2,-1/2,0),  \\
u_R &\sim (3,1,2/3,r), &   \hspace{-0.2 cm} e_R &\sim (1,1,-1,-r),  \\
d_R &\sim (3,1,-1/3,-r), &  \hspace{-0.2 cm} \nu_R &\sim (1,1,0,r), 
\end{aligned}
\end{equation}
is anomaly free under ${\cal G}_R$. The SM Higgs, $H\sim (1,2,1/2,r)$, is also charged under $\text{U}(1)_R$ in order to give mass to the fermions through the usual Yukawa interactions,
\begin{equation}
\begin{split}
-{\cal L} \supset & \ Y_u \, \bar q_L \tilde H u_R + Y_d \, \bar q_L H d_R \\
& + Y_\nu \, \bar \ell_L \tilde H \nu_R + Y_e \, \bar \ell_L H e_R + \text{h.c.},
\end{split}
\end{equation}
where $\tilde H = i \sigma_2 H^*$.
To implement the NB mechanism, let us consider extra vector-like fermions that under ${\cal G}_R$ transform as\footnote{We call this new abelian gauge symmetry ${\rm U}(1)_R$ because that is what it is acting on the SM fermions. For the new vector-like fermions the left handed fields also have ${\rm U}(1)_R$ charge.}:
\begin{equation}
\begin{aligned}
U_R &\sim (3,1,2/3,-R),  & U_L &\sim (3,1,2/3,L),\\
D_R & \sim (3,1,-1/3,R), & D_L &\sim (3,1,-1/3,-L),\\
E_R & \sim (1,1,-1,R), & E_L &\sim (1,1,-1,-L), \\
N_R & \sim (1,1,0,-R), & N_L &\sim (1,1,0,L).
\end{aligned}
\end{equation}
 This set of vector-like under the SM\footnote{Henceforth we will just use {\it vector-like} to refer to these fermions.} fermions is anomaly free for any value of the charges $L$ and $R$. 
The new fermions get mass through the spontaneous breaking of $\text{U}(1)_R$ by the vacuum expectation value (VEV) of the scalar $S \sim (1,1,0,L+R)$, via the following interactions, 
\begin{equation}
\begin{split}
-{\cal L} \supset &\left(\lambda_U \bar U_L U_R +\lambda_N \bar N_L N_R \right) S \\
&+ \left( \lambda_D \bar D_L  D_R + \lambda_E \bar E_L  E_R  \right) S^*+ \text{h.c.},
\end{split}
\end{equation}
where we have assumed $L\neq -R$.
For generic $\text{U}(1)_{\rm R}$ charges, all the fermions (including the neutrinos) have Dirac masses, and baryon and lepton number are conserved at the renormalizable level.

At this point the vector-like fermions are stable. In order to implement the NB mechanism~\cite{Nelson:1983zb,Barr:1984qx} and allow the vector-like fermions to decay, 
new scalar fields are needed. These scalars trigger SCPV through a complex VEV.  There are  two complex scalar fields $X_a \sim (1,1,0,L-r)$ with $a=1,2$.
The $\text{U}(1)_R$ charge of $X_a$ fields allows the following interaction terms in the Lagrangian,
\begin{equation}
\begin{split}
\label{eq:XYuk}
&-{\cal L}  \supset   \bar U_L u_R^i \! \left(   \sum_{a=1}^{2}  \lambda_{u,a}^{i} X_a \! \right) \! + \! \bar D_L d_R^i \! \left(  \sum_{a=1}^{2} \lambda_{d,a}^{i} X_a^* \! \right) \\
& \! + \! \bar N_L \nu_R^i \! \left(  \sum_{a=1}^{2} \lambda_{\nu,a}^{i} X_a \! \right) \! + \! \bar E_L e_R^i \!  \left( \sum_{a=1}^{2} \lambda_{e,a}^{i} X_a^*\! \right) \! + \! \text{h.c.}.
\end{split}
\end{equation}
Note that $r\neq R$, otherwise the Higgs doublet can couple to the right-handed new fermions, spoiling the Nelson-Barr mechanism.

Because the Higgs boson is charged under $\text{U}(1)_{\rm R}$ in this model, there is a tree-level mixing between the electroweak neutral gauge boson, $Z$, and the abelian generator of $\text{U}(1)_{\rm R}$, $Z_R$, in the broken phase:
\begin{equation}
{\cal L} \supset \frac{1}{2}\begin{pmatrix} Z^\mu & Z_R^\mu \end{pmatrix} \begin{pmatrix} M_Z^2 & - \frac{2g_R r  M_Z^2}{\sqrt{g_1^2 + g_2^2}} \\ - \frac{2g_R r M_Z^2}{\sqrt{g_1^2+g_2^2}} & M_{Z_R}^2\end{pmatrix} \begin{pmatrix} Z_\mu \\ Z_{R\mu} \end{pmatrix}.
\end{equation}
In the following we neglect the mixing angle between $Z$ and $Z_R$, given by $2 g_R r M_Z^2 /(\sqrt{g_1^2+ g_2^2}\,  M_{Z_R}^2)$,
since $M_{Z_R} > M_{Z}$ and $g_R \ll 1$ to ensure an acceptably small correction to the Z mass and small enough flavor changing neutral currents.

Although this paper focuses on ${\rm U}(1)_R$, it is worth noting that one could change the new gauge group from ${\rm U}(1)_R$ to ${\rm U}(1)_{X}$, as long as the SM fermions are anomaly free under the new gauge symmetry, with $L$ and $R$ now interpreted as $X$ charges. For example, $X$ could be $B-L$. For the ${\rm U}(1)_{B-L}$ case, the $X_i$'s cannot simultaneously couple to quarks and leptons because they have different $B-L$ charges. This will lead to charged stable relics\footnote{Stable relics are not necessarily a problem if inflation occurs at a scale lower than their mass.} except for the case when $L=1/3$. In this case, $X_i\sim (1,1,0,2/3)$ couple to both the charged leptons and the down-type quarks, while the vector-like up-type quarks are connected to the SM through the mass term $\bar U_L u_R$. Lepton number is violated by the interaction $N_L^T C N_L X_i^*$. The vector-like neutrino, however, is stable and therefore could be a candidate for cold dark matter~\cite{FileviezPerez:2017sbb}.

\section{SCPV and The Strong CP-problem}

Working in the basis where the VEV of $S$ is made real via a ${\rm U}(1)_R$ transformation, one can generate a vacuum expectation value with a non-zero phase for at least one of the $X_i$ fields using the scalar potential,
\begin{equation}
V \supset - \mu_X^2 (X_1^\dagger X_2 ) + \lambda_X (X_1^\dagger X_2)^2 + \text{h.c.}.
\end{equation}  
Defining the VEVs as $\left<X_a \right>= e^{i \theta_a} v_{X_a} / \sqrt{2}$, one finds
\begin{equation}
V\supset - \mu_X^2 v_{X_1} v_{X_2} \cos \theta_X + \lambda_X \frac{v_{X_1}^2 v_{X_2}^2}{2} \cos 2 \theta_X,
\end{equation}
with $\theta_X=\theta_2 - \theta_1$. Here, $\mu_X$ is an ``effective mass" containing all possible contributions to the term $X_1^\dagger X_2$ ({\it e.g.} $X_1^\dagger X_1$).
The minimization condition, $\partial V/\partial \theta_X =0$, gives us 
\begin{equation}
\cos \theta_X =  \frac{\mu_X^2}{2 \lambda_X v_{X_1} v_{X_2} }.
\label{eq:costhetaX}
\end{equation}
Even in the limit of large VEVs, $\cos \theta_X$ can still be order one as long as $\mu_X \sim v_{X}$. Only the combination of phases $\theta_X$ is determined by the scalar potential. To determine the remaining free phase, particular values of the charges are required which allow for additional terms.\footnote{We thank Lisa Randall for pointing this out to us.} For example, $2L = 3 r + R$ allows $S^*X_i X_j X_k$ terms. Note that adding these terms will modify Eq.~\eqref{eq:costhetaX}. These terms explicitly break the ${\rm U}(1)_{X_1 - X_2}$ global symmetry that otherwise would lead to a Nambu-Goldstone boson.

In the broken phase, the fermion masses are given by
\begin{equation}
-  {\cal L} \supset \bar f_L'^{A} {\cal M}_f^{AB}  f_R'^B + \text{h.c.}, 
\end{equation}
with the mass matrices at tree-level given by
\begin{equation}
{\cal M}_f^{AB} = \frac{1}{\sqrt{2}} \begin{pmatrix} \displaystyle Y_f^{ij} v_H  && 0^{i4} \\ \displaystyle \sum_{a=1}^2 \lambda_{f,a}^j v_{X_a} e^{\pm i\theta_a} && \displaystyle \lambda_F   v_S \end{pmatrix},
\label{eq:Mf}
\end{equation}
where $\langle S \rangle = v_S / \sqrt{2}$. In our convention, capital letters $A,B,... = 1,2,3,4$, roman letters $i,j,...=1,2,3$ run over the indices of the light quarks and $4$ will refer to the new vector-like fermions. In the above matrix when $A$ goes over 1,2,3 is represented by $i$ and when $B$ goes over 1,2,3 is represented by $j$. Above, the primes are used for the weak eigenstates. In Eq.~\eqref{eq:Mf}, $\pm$ takes $+$ ($-$) for $U$ and $N$ ($E$ and $D$) fermions.

We have used the ${\rm U}(1)_R$ gauge symmetry to set the phase of $\langle S \rangle$ to be zero. However, if we had not done that this phase cancels out in ${\rm arg} \{ \rm Det ({\cal M}_u {\cal M}_d)\}$ because $S$ gives mass to $U$'s while $S^*$ gives mass to $D$'s. 

Both a CKM phase in the quark sector, and a Pontecorvo-Maki-Nakagawa-Sakata (PMNS) phase in the lepton sector are generated from the same phase $\theta_X$.

In this model $\theta_{\rm QCD}=0$ by CP invariance and the argument of the determinant of the mass matrices can be seen to be zero by expanding these determinants in minors about the last column. Hence in this model the renormalizable couplings give  $\bar \theta_{\rm QCD}=0$  at tree level.

\subsection{Non-renormalizable operators and loops}

Non-renormalizable operators and radiative corrections can give rise to CP violating corrections to the mass matrices  (see Refs.~\cite{Bento:1991ez,Asadi:2022cco}), $\delta {\cal M}_q$, giving rise to the following correction to $\bar \theta_{\rm QCD}$,
\begin{equation}
\Delta \bar \theta_{\rm QCD} \simeq \text{Im} \big \{\text{Tr}\{({\cal M}_q^{-1} \delta {\cal M}_q\} \big \}.
\label{eq:thetaQCDcorr}
\end{equation}
\begin{figure}
\begin{equation*}
\begin{gathered}
\begin{tikzpicture}[line width=1.5 pt,node distance=1 cm and 1 cm]
\coordinate[label=left:$t_R$](dR);
\coordinate[right = 1 cm of dR](v1);
\coordinate[right = 2 cm of v1](v2);
\coordinate[right= 1 cm of v2,label=right:$t_L$](dL);
\coordinate[right= 1 cm of v1](vaux);
\coordinate[below =  0.75 cm of vaux,label=below:$\langle X_k \rangle$](Xk);
\coordinate[above = 1 cm of vaux](v4);
\coordinate[above left = 1 cm of v4,label=left:$\langle X_i \rangle$](Xi);
\coordinate[above right = 1 cm of v4,label=right:$\langle H \rangle$](Xj);
\coordinate[right = 0.5cm of v1,label=below:$U_L$](DLlabel);
\coordinate[right = 0.5cm of vaux,label=below:$t_R$](dRlabel);
\coordinate[above = 0.75 cm of v1,label=left:$X_\ell$](Xlabel);
\coordinate[above = 0.75 cm of v2,label=right:$H$](Xlabel);
\draw[scalarnoarrow] (v4)--(Xi);
\draw[scalarnoarrow] (v4)--(Xj);
\draw[fermion](dR)--(v1);
\draw[fermion](v1)--(vaux);
\draw[fermion](vaux)--(v2);
\draw[fermion](v2)--(dL);
\draw[scalarnoarrow](vaux)--(Xk);
\draw[fill=black] (v1) circle (.05cm);
\draw[fill=black] (v2) circle (.05cm);
\draw[fill=black] (vaux) circle (.05cm);
\draw[fill=black] (v4) circle (.05cm);
\semiloop[scalarloop]{v1}{v2}{0};
\semiloop[scalarloop1]{v1}{v2}{0};
\end{tikzpicture}
\end{gathered}
\end{equation*}

\begin{equation*}
\begin{gathered}
\begin{tikzpicture}[line width=1.5 pt,node distance=1 cm and 1 cm]
\coordinate[label=left:$U_R$](DR);
\coordinate[right = 0.75 cm of DR](vS);
\coordinate[right = 0.5 cm of vS,label=below:$U_L$](labelDL);
\coordinate[above = 0.75 cm of vS,label=above:$\langle S \rangle$](S);
\coordinate[right = 1 cm of vS](v1);
\coordinate[right = 2 cm of v1](v2);
\coordinate[right= 0.75 cm of v2,label=right:$t_L$](dL);
\coordinate[right= 1 cm of v1,label=below:$t_{R}$](vaux);
\coordinate[above = 1 cm of vaux](v4);
\coordinate[above left = 1 cm of v4,label=left:$\langle X_i \rangle$](Xi);
\coordinate[above right = 1 cm of v4,label=right:$\langle H \rangle$](Xj);
\coordinate[above = 0.75 cm of v1,label=left:$X_k$](Xlabel);
\coordinate[above = 0.75 cm of v2,label=right:$H$](Xlabel);
\draw[scalarnoarrow] (v4)--(Xi);
\draw[scalarnoarrow] (v4)--(Xj);
\draw[fermion](dR)--(vS);
\draw[fermion](vS)--(v1);
\draw[fermion](v1)--(v2);
\draw[fermion](v2)--(dL);
\draw[scalarnoarrow](vS)--(S);
\draw[fill=black] (vS) circle (.05cm);
\draw[fill=black] (v1) circle (.05cm);
\draw[fill=black] (v2) circle (.05cm);
\draw[fill=black] (v4) circle (.05cm);
\semiloop[scalarloop]{v1}{v2}{0};
\semiloop[scalarloop1]{v1}{v2}{0};
\end{tikzpicture}
\end{gathered}
\end{equation*}

\caption{Some 1-loop diagrams contributing to the fermion mass matrices that give a non-zero $\bar \theta_{\rm QCD}$.}
\label{fig:deltaMqq}
\end{figure}
Assuming an order one CP violating phase, the 1-loop diagrams in Fig.~\ref{fig:deltaMqq} give 
\begin{equation}\label{eq:1loop}
\Delta \bar \theta_{\rm QCD} \sim \frac{\lambda}{16\pi^2} \left(\frac{\tilde M_U}{m_X}\right)^2 \!\! \text{ln}\left(\frac{m_X^2}{m_t^2}\right),
\end{equation}
where $\lambda$ is the $H^\dagger H X_i^* X_j$ quartic coupling, and $\tilde M_U$ is the mass of the up-type new vector-like quark.
One way for this to be consistent with the experimental bounds is to have $m_X \sim \tilde M_U$ and $\lambda$ very small (which does not require fine tuning). Another way is to have $\lambda$ of order one and $\tilde M_U \ll m_X$. For example, if $m_X \sim 10^{11}$ GeV and the mass of the new vector-like fermions is $M_U \sim 5$ TeV, $\lambda$ is not constrained by the neutron EDM.

Contributions to the vector-like fermion mass terms from higher dimension operators can spoil the NB solution to the strong CP puzzle. For example,
\begin{equation}\label{eq:higherdimop}
\frac{1}{\Lambda_{\rm EFT}^2} \bar U_L U_R \, S \, X_1 X_2^*  +\frac{1}{\Lambda_{\rm EFT}^2} \bar q_L \tilde H u_R \,  X_1 X_2^*  +  \cdots.
\end{equation}
Note that if the vector-like fermion has a bare mass term then there would be dimension five operators that spoil the NB mechanism, which might be problematic given cosmological constraints on completely stable domain walls arising from SCP~\cite{McNamara:2022lrw}. Such a mass term is forbidden if $L \neq R$.
According to Eq.~\eqref{eq:thetaQCDcorr}, the higher dimensional operators explicitly displayed in Eq.~\eqref{eq:higherdimop} shift the $\bar \theta_{\rm QCD}$ as follows
\begin{equation}
\Delta \bar \theta_{\rm QCD} \sim  \frac{1}{\lambda_F}\frac{v_X^2}{\Lambda_{\rm EFT}^2} + \frac{1}{Y_u} \frac{v_X^2}{\Lambda_{\rm EFT}^2},
\end{equation}
where we assumed order one CP violating phases. This leads to the following condition
\begin{equation}
\label{eq:SCPbound}
v_X \lesssim 10^{14} \text{ GeV }\left(\frac{\Lambda_{\rm EFT}}{M_{\rm Pl}}\right) / \sqrt{\frac{1}{\lambda_F}+\frac{1}{Y_u}}.
\end{equation}
Note that the upper bound above cannot surpass $v_X \lesssim 10^{11} \text{ GeV}$ since $Y_u \sim m_u/v_H$. This is high enough for inflation to  have occurred after the spontaneous breaking of CP, but too low for the gravitational waves from inflation~\cite{Rubakov:1982df,Fabbri:1983us,Abbott:1984fp,Starobinsky:1985ww} to be observable in the  B modes of the CMB~\cite{Kamionkowski:1996zd,Kamionkowski:1996ks,Seljak:1996gy,Zaldarriaga:1996xe}.


%
\section{The CKM and PMNS matrices}
%
We start by discussing how the CKM and PMNS phases arise in Nelson-Barr models. 
The $4 \times 4$ fermion mass matrix in Eq.~\eqref{eq:Mf} is generic to those models when there is only one generation of new vector-like fermions.
Let us redefine the fermions as follows
\begin{equation}
f'_L = O_{\textsc{f}L} f_L'', \quad \text{and} \quad f_R' = O_{\textsc{f}R} f_R'',
\end{equation}
where ${\cal O}_{\textsc{f}L,R}$ is an orthogonal matrix, 
\begin{equation}
O_{\textsc{f}L,R}^{AB} = \begin{pmatrix} & & & 0 \\ & o_{\textsc{f}L,R}^{ij} & & 0 \\ & & & 0  \\ 0 & 0 & 0 & 1 \end{pmatrix},
\end{equation}
that diagonalizes the light quark masses $3 \times 3$ block\footnote{We note that $m_i$ are not the mass eigenstates of the light quarks, as the total $4 \times 4$ matrix is not diagonalized yet. However, as we will show later, they are expected to be of the same order.},
\begin{equation}
\left(o_{\textsc{f}L}^T Y_f o_{\textsc{f}R}\right) \frac{v_H}{\sqrt{2}} =  \text{diag}(m_1,m_2,m_3) .
\end{equation}
Thus,
\begin{equation}
\begin{split}
-{\cal L} &= \bar f_L'' \, O_{\textsc{f}L}^T \, {\cal M}_f \, O_{\textsc{f}R} \, f_R'' + \text{h.c.}, \\
&=\bar f_L'' \begin{pmatrix} m_1 & 0 & 0  & 0 \\ 0 & m_2 & 0 & 0 \\ 0 & 0 & m_3 & 0 \\ \mu_1 & \mu_2 & \mu_3 & M_F \end{pmatrix} f_R'' + \text{h.c.},
\end{split}
\label{eq:mOr}
\end{equation}
 where 
 \begin{equation}
     \mu_j = \sum_k  \sum_{a=1}^2 \lambda_{f,a}^k v_{X_a} e^{\pm i\theta_a}  \, o_{\textsc{f}R}^{kj}. 
 \end{equation}

Further transformations on the left handed and right handed fields are required to fully diagonalize the fermion mass matrix. 

Diagonalizing the hermitian matrix ${\cal M}_f {\cal M}_f^\dagger$ determines the CKM matrix (and the PMNS matrix). We will denote the matrix that diagonalizes ${\cal M}_f {\cal M}_f^\dagger$ by $\tilde V_{\textsc{f}L}$. Starting from the matrix in Eq.~\eqref{eq:mOr}, we find that the hermitian matrix ${\cal M}_f{\cal M}_f^\dagger$ in the double primed basis is given by
\begin{equation}
    O_{\textsc{f}L}^T {\cal M}_f {\cal M}_f^\dagger O_{\textsc{f}L} = \begin{pmatrix} m_1^2 & 0 & 0 & \mu_1^* m_1 \\ 0 & m_2^2 & 0 & \mu_2^* m_2 \\ 0 & 0 & m_3^2 & \mu_3^* m_3 \\ \mu_1 m_1 & \mu_2 m_2 & \mu_3 m_3 & \tilde M_F^2 \end{pmatrix},
    \label{eq:MMdagger}
\end{equation}
where $\tilde M_F^2 = M_F^2 + \sum_i |\mu_{i}|^2$. The matrix 
$V_{\textsc{f}L}$ is given approximately by
\begin{equation}
V_{\textsc{f}L} = \begin{pmatrix} 1 & 0 & 0 & \tfrac{m_1 \mu_1^*}{\tilde M_F^2} \\ 0 & 1 & 0 & \tfrac{m_2 \mu_2^*}{\tilde M_F^2} \\
0 & 0 & 1 & \tfrac{m_3 \mu_3^*}{\tilde M_F^2} \\
- \tfrac{m_1 \mu_1}{\tilde M_F^2} & - \tfrac{m_2 \mu_2}{\tilde M_F^2} & -\tfrac{m_3 \mu_3}{\tilde M_F^2}& 1 \end{pmatrix}.
\label{eq:VL}
\end{equation}

\vspace{0.2cm}

The non-zero off-diagonal elements are correct up to ${\cal O}(m^2 \mu^2 / \tilde M_F^4)$ and $V_{\textsc{f}L}^\dagger V_{\textsc{f}L} = \mathbb{I} + {\cal O}(m^2 \mu^2 / \tilde M_F^4)$.  The matrix in Eq.~\eqref{eq:VL}
approximately diagonalizes Eq.~\eqref{eq:MMdagger}, leading to the following matrix,
\begin{widetext}
\begin{equation}
V_{\textsc{f}L}^\dagger O_{\textsc{f}L}^T {\cal M}_f {\cal M}_f^\dagger O_{\textsc{f}L} V_{\textsc{f}L} =
    \begin{pmatrix}
    m_1^2\left(1 - \frac{|\mu_1|^2}{\tilde M_F^2}\right) & - m_1 m_2 \frac{\mu_1^*\mu_2}{\tilde M_F^2} & -m_1 m_3 \frac{\mu_1^* \mu_3}{\tilde M_F^2} &  {\cal O}\! \left(\tfrac{m^3 \mu}{\tilde M_F^2}\right) \\
    -m_2 m_1 \frac{\mu_2^* \mu_1}{\tilde M_F^2} & m_2^2\left(1  - \frac{|\mu_2|^2}{\tilde M_F^2}\right) & -m_2 m_3 \frac{\mu_2^* \mu_3}{\tilde M_F^2} & {\cal O} \! \left(\tfrac{m^3\mu}{\tilde M_F^2}\right) \\
    -m_3 m_1 \frac{\mu_3^*\mu_1}{\tilde M_F^2} & - m_3 m_2 \frac{\mu_3^*\mu_2}{\tilde M_F^2} & m_3^2 \left(1 - \frac{|\mu_3|^2}{\tilde M_F^2}\right) & {\cal O} \! \left(\tfrac{m^3 \mu}{\tilde M_F^2}\right) \\
     {\cal O} \! \left(\tfrac{m^3 \mu}{\tilde M_F^2}\right) & {\cal O}\! \left(\tfrac{m^3 \mu}{\tilde M_F^2}\right) &  {\cal O} \! \left(\tfrac{m^3\mu}{\tilde M_F^2}\right) & \tilde M_F^2 + 2\sum_i m_{i}^2 \frac{ |\mu_{i}|^2 }{\tilde M_F^2} + {\cal O} \! \left(\tfrac{m^4}{\tilde M_F^4}\right)
    \end{pmatrix}.
\end{equation}
\end{widetext}
The off-diagonal elements in the 4th row and column are very suppressed. Therefore, the CKM matrix comes mostly from diagonalizing the $3\times 3$ upper left block. Furthermore, if $\mu / \tilde M_F \ll 1$, then the upper left $3 \times 3$ block is approximately diagonal and the CKM phase is small, as already noted in Refs.~\cite{Vecchi:2014hpa,Valenti:2021rdu}.

Since $\mu/\tilde M_F$ is not small, the $3\times 3$ upper-left block of the above matrix still needs to be diagonalized. As it is hermitian, it will be diagonalized by a unitary (complex) matrix,
\begin{equation}
U_{\textsc{f}L}^{AB} = \begin{pmatrix} & & & 0 \\ & \textsc{u}_{\textsc{f}L}^{ij} & & 0 \\ & & & 0 \\ 0 & 0 & 0 & 1 \end{pmatrix}.
\end{equation}
The unitary matrix that diagonalizes ${\cal M}_f{\cal M}_f^\dagger$ is then
\begin{equation}
\tilde V_{\textsc{f}L} = O_{\textsc{f}L} V_{\textsc{f}L} U_{\textsc{f}L},
\label{eq:VLtilde}
\end{equation}
and therefore the CKM matrix is given by following matrix,
\begin{eqnarray}
\label{eq:VCKM}
 &&V_\text{CKM}^{ij} = \sum_k (\tilde V_{\textsc{U}L}^\dagger)^{ik} (\tilde V_{\textsc{D}L})^{kj} \\
&&=\sum_{k,\ell} (\textsc{u}_{\textsc{u}L}^\dagger)^{ik}  \left(o_{\textsc{u}L}^To_{\textsc{d}L}\right)^{k\ell}  \textsc{u}_{\textsc{d}L}^{\ell j}
+ {\cal O}\left(\frac{m^2 \mu^2}{\tilde M_U^2 \tilde M_D^2}\right),
\nonumber
\end{eqnarray}
which agrees with the previous literature~\cite{Branco:2003rt}.

In these models, CP violation in the lepton sector arises in the same fashion as in the quark sector and a CP violating phase in the PMNS matrix is expected. However, since it is unconstrained by experiment, in the lepton sector $\mu$ could be much smaller than $\tilde M_F$.

%
\section{Flavor Violation}
%
In this theory we have new flavor violating interaction due to mixing between the SM fermions with the new heavy fermions.

\subsection{Z boson couplings}
Flavor changing neutral currents only enter through the left-handed fermions, and are suppressed by ${\cal O}(m^2 \mu^2 / \tilde M_F^4)$,

\pagebreak

\begin{eqnarray}\label{eq:FCNC}
&&{\cal L} \supset -\frac{2e \,T_3^{f_L} }{\sin 2\theta_W }    Z^\mu\\
&& \ \  \times \bar f_{L}^i \gamma_\mu \left( \sum_{\ell,k} (\textsc{u}_{\textsc{f}L}^\dagger)^{ik} \frac{m_k m_\ell \mu_k^* \mu_\ell}{\tilde M_F^4} (\textsc{u}_{\textsc{f}L})^{\ell j} \right)  f_{L}^j,  \nonumber
\end{eqnarray}
where $T_3^{f_L} = \pm 1/2$ is the weak isospin of the left-handed fermion $f_L$.
Even though Eq.~\eqref{eq:FCNC} is of ${\cal O}(m^2 \mu^2/\tilde M_F^4)$, which is of the same order as the error in the elements of the matrix in Eq.~\eqref{eq:VL}, the interaction above arises from the product of two entries that are of ${\cal O}(m \, \mu / \tilde M_F^2)$. These are much more suppressed that the FCNC in Ref.~\cite{Ishiwata:2015cga}.
Eq.~\eqref{eq:FCNC} agrees with the results of Ref.~\cite{Perez:2020dbw}. 

The coupling of the heavy vector-like fermions to the SM fermions and the $Z$,
\begin{eqnarray}
&&{\cal L} \supset \frac{2e \,  T_3^{f_L}}{\sin 2\theta_W}  Z^\mu\\
&& \qquad \times \bar f_{L}^i \gamma_\mu\left( \sum_k (\textsc{u}_{\textsc{f}L}^\dagger)^{ik} \frac{m_k \mu_k^*}{\tilde M_F^2} \right)  f_{L}^4 + \text{h.c.}. \nonumber 
\end{eqnarray}
Since in this theory the new fermions are heavy, bounds from flavor violating processes can be satisfied. For example, let us consider the stringent constraint on the $\mu$ to $e$ conversion in nuclei. Given the Feynman rule for the $Z$ boson mediating flavor violating interactions with electrons and muons from Eq.~\eqref{eq:FCNC}, the limit on the conversion of muons to electrons in gold~\cite{SINDRUMII:2006dvw} requires~\cite{deGouvea:2013zba,Bernabeu:1993ta},
\begin{equation}
    \left(\frac{m_\ell}{\tilde M_E}\right) < 5 \times 10^{-4} \left(\frac{{\cal B}(\mu \to e) \text{ [Au]}}{7 \times 10^{-13}}\right)^{1/4} \left(\frac{\tilde M_E/\mu_\ell}{1}\right).
\end{equation}
For $m_\ell \sim m_\tau$, the above constrain is satisfied if $\tilde M_E > 4$ TeV. Projected bounds on this process~\cite{Dornan:2016zsy} are about four orders of magnitude stronger and will increase the bound on $\tilde M_E$ by an order of magnitude. A more comprehensive analysis may be warranted.

\subsection{W boson couplings}
The couplings of the heavy vector-like fermions with the $W$ bosons and the SM fermions are given by
\begin{widetext}
\begin{eqnarray}
    {\cal L} \supset && \frac{g_2}{\sqrt{2}}W_\mu^+ \\
    && \times \sum_{k,p,\ell} \left(  \bar u_{L}^i(\textsc{u}_{\textsc{u}L}^\dagger)^{i \ell} (\textsc{o}_{\textsc{u}L}^T)^{\ell p} (\textsc{o}_{\textsc{d}L})^{p k} \frac{m_{\textsc{d}k} \mu_{\textsc{d}k}^*}{\tilde M_D^2}  \gamma^\mu d_{L}^4 + \bar u_{L}^4 \frac{m_{\textsc{u}k}\mu_{\textsc{u}k}}{\tilde M_U^2}(\textsc{o}_{\textsc{u}L}^T)^{kp} (\textsc{o}_{\textsc{d}L})^{p\ell} (\textsc{u}_{\textsc{d}L})^{\ell i} \gamma^\mu d_{L}^i \right) + \text{h.c.}. \nonumber
\end{eqnarray}
\end{widetext}
We specify the involved fermion in the label of $m$ and $\mu$ to note which mass matrix they come from.

The SM fermion couplings with the $W$ boson are given by the CKM matrix, explicitly given in Eq.~\eqref{eq:VCKM}. 

Similar formulas hold for the leptons. 

\subsection{SM-Higgs couplings}
The couplings of the light fermions to the SM Higgs are given by the following Yukawa interaction,
\begin{equation}
-{\cal L} \supset \left(\frac{h}{v_H}\right) \bar f_{L}'^{i} {\cal M}_f^{ij} f_{R}'^{j} + \text{h.c.},
\label{eq:HiggsLag}
\end{equation}
where the primes indicate that the fermions are weak eigenstates. Using the following identity,
\begin{equation}
\label{eq:33}
\begin{split}
& \sum_{\ell,k} (\tilde V_{\textsc{f}L}^\dagger)^{i\ell} {\cal M}_f^{\ell k} (\tilde V_{\textsc{f} R})^{kj}\\
& \  = (\tilde V_{\textsc{f}L}^\dagger {\cal M}_f \tilde V_{\textsc{f}R})^{ij} - \sum_A (\tilde V_{\textsc{f}L}^\dagger)^{i4} {\cal M}_f^{4A} (\tilde V_{\textsc{f}R})^{Aj}.
\end{split}
\end{equation}
where $\tilde V_{\textsc{f}R}$ diagonalizes ${\cal M}_f^\dagger {\cal M}_f$, we can rewrite the above Lagrangian in the mass eigenstate basis as follows,
\begin{equation}
-{\cal L} \supset \frac{h}{v_H} \bar f_{L}^i \left( \tilde m_i \delta^{ij} -\sum_A   (\tilde V_{\textsc{f}L}^\dagger)^{i4} {\cal M}_f^{4A} \tilde V_{\textsc{f}R}^{Aj}\right) f_{R}^j + \text{h.c.},
\end{equation}
where $\tilde m_i$ are the physical light fermion masses.
The first part, which corresponds to the diagonal matrix $\tilde m_i \delta^{ij}$, is the standard model coupling to the Higgs boson. The second term will lead to flavor changing interactions which are CP violating. We can use the identity $({\cal M}_f \tilde V_{\textsc{f}R})^{4j} = (\tilde V_{\textsc{f}L} {\cal M}_\text{diag})^{4j}$ to rewrite the second term above as, 
\begin{equation}
-{\cal L} \supset -\frac{h}{v_H} \bar f_{L}^i (\tilde V_{\textsc{f}L}^\dagger)^{i4} ( \tilde  V_{\textsc{f}L} {\cal M}_\text{diag})^{4j} f_{R}^j + \text{h.c.},
\end{equation}
where ${\cal M}_\text{diag} = \text{diag}(\tilde m_1, \tilde m_2, \tilde m_3, \tilde M_F)$ is the diagonal matrix with the physical masses of the fermions. Expressing $\tilde V_{\textsc{f}L}$ as in Eq.~\eqref{eq:VLtilde}, and exploiting the fact that $(U_{\textsc{f}L})^{A4} =(U_{\textsc{f}L})^{4A} = \delta^{A4}$ and $(O_{\textsc{f}L})^{A4} =(O_{\textsc{f}L})^{4A} = \delta^{A4}$, and that $(V_{\textsc{f}L})^{4i} = - m_i \mu_i / \tilde M_F^2$, we can write the non-CP conserving light fermion couplings to the standard model Higgs boson as,
\begin{eqnarray}
&&-{\cal L} \supset -h \left( \frac{\tilde m_j}{v_H} \right) \bar f_{L}^i \sum_{\ell,k} (\textsc{u}_{\textsc{f}L}^\dagger)^{i\ell} \frac{m_\ell \mu_\ell^* m_k \mu_k}{\tilde M_F^4}(\textsc{u}_{\textsc{f}L})^{kj}  f_{R}^j \nonumber\\
&& \qquad + \text{h.c.}.
\end{eqnarray}

Similarly, we can write the coupling between the new fermions, the standard model fermions and the Higgs boson in the following way,
\begin{equation}
\label{eq:Higgs4i}
- {\cal L} \supset  \bar f_{L}^i \sum_k (\textsc{u}_{\textsc{f}L}^\dagger)^{ik} \frac{\mu_k^*}{\tilde M_F} \left(\frac{m_k}{v_H}\right) h f_{R}^4 + \text{h.c.},
\end{equation}
which will be relevant for the decay rates of the heavy vector-like fermions. Note that for $\mu \sim \tilde M_F$ the amplitude for the interaction that allows $f_4$ to decay to a Higgs and a SM fermion is not suppressed by $\tilde M_F$.

Even without a detailed discussion of the couplings of the $X$ and $S$ scalars and $Z_R$ gauge boson to the fermions, upper bounds can be derived on the lifetimes of the new vector-like fermions. These upper bounds are dominated by decays through the Higgs boson via the couplings in Eq.~\eqref{eq:Higgs4i},
\begin{equation}
\begin{split}
\tau_F  & < \left[\frac{\tilde M_F}{8\pi}  \left(\frac{\mu_F}{\tilde M_F}\right)^2\left(\frac{m_f}{v_H}\right)^2\right]^{-1},
\end{split}
\end{equation}
where $m_f$ is the mass of the heaviest fermion of its kind (electrons, neutrinos, up-type or down-type quarks).

The measured CKM phase implies $\mu_Q \sim \tilde M_Q$, which fixes the upper bound on the  lifetime of the new quarks to be around $10^{-21} \, {\rm s}$ ($10^{-26 } \, {\rm s}$) for TeV down (up) vector-like quark masses. On the other hand, the PMNS phase in the lepton sector is unknown, which renders more freedom to the hierarchy between $\mu$ and $\tilde M_F$. Particularly interesting is the case of the vector-like neutrino, $N$, as the upper bound on its lifetime is strongly suppressed by the SM neutrino masses:
\begin{equation}
\begin{split}
\tau_N 
& < 0.1 \text{ s} \left(\frac{1 \text{ TeV}}{\tilde M_N}\right) \! \left(\frac{\tilde M_N / \mu_N}{1}\right)^2 \! \left( \frac{0.1 \text{ eV}}{m_\nu}\right)^2.
\end{split}
\end{equation}
If the PMNS phase turns out to be of the same magnitude or bigger than the CKM phase, then $\mu_{\textsc{n}} \sim \tilde M_N$ and a TeV scale vector-like neutrino would be expected to decay before BBN.

\section{Finite Naturalness}
%
\begin{figure}[t!]
\begin{equation*}
\begin{gathered}
\begin{tikzpicture}[line width=1.5 pt,node distance=1 cm and 1 cm]
\coordinate[label=left:$H$] (H1);
\coordinate[right = 0.75 cm of H1](v1);
\coordinate[right = 0.75 cm of v1,label=right:$H$](H2);
\draw[scalar] (H1)--(v1);
\draw[scalar] (v1)--(H2);
\draw[fill=black] (v1) circle (.05cm);
\end{tikzpicture}
\end{gathered} \ + \  \ \
\begin{gathered}
\begin{tikzpicture}[line width=1.5 pt,node distance=1 cm and 1 cm]
\coordinate[] (g1);
\coordinate[below = 0.85 cm of g1](g2);
\coordinate[right = 1 cm of g1](g3);
\coordinate[right = 1 cm of g2](g4);
\draw[gluon](g1)--(g3);
\draw[gluon](g2)--(g4);
\draw[fermion](-0.7,-0.4)  circle (0.8);
\draw[fermionp](1.7,-0.4)  circle (0.8);
\coordinate[right = 1.4 cm of g3](H1);
\coordinate[right = 1.4 cm of g4](H2);
\coordinate[above right = 0.75 cm of H1, label=right:$H$](H1f);
\coordinate[below right = 0.75 cm of H2, label=right:$H$](H2f);
\draw[fill=black] (H1) circle (.05cm);
\draw[fill=black] (H2) circle (.05cm);
\draw[scalar] (H1)--(H1f);
\draw[scalar] (H2f)--(H2);
\coordinate[below = 0.6 cm of g1](gaux);
\coordinate[right = 1.7 cm of gaux,label=below:$t$](top);
\coordinate[left = 0.7 cm of gaux,label=below:$F$](F);
\end{tikzpicture}
\end{gathered}
\end{equation*}
\begin{equation*}
\begin{gathered}
\begin{tikzpicture}[line width=1.5 pt,node distance=1 cm and 1 cm]
\coordinate[label=left:$H$] (i1);
\coordinate[below right = 1.5 cm of i1](v1);
\coordinate[below left = 1.5 cm of v1,label=left:$X_a$](i2);
\coordinate[above right = 1.5 cm of v1,label=right:$H$](i3);
\coordinate[below right = 1.5 cm of v1,label=right:$X_b$](i4);
\draw[scalar](i1)--(v1);
\draw[scalar](v1)--(i2);
\draw[scalar](v1)--(i3);
\draw[scalar](i4)--(v1);
\draw[fill=black] (v1) circle (.05cm);
\end{tikzpicture}
\end{gathered}  \ \  + \ \   
\begin{gathered}
\begin{tikzpicture}[line width=1.5 pt,node distance=1 cm and 1 cm]
\coordinate[label=left:$H$] (i1);
\coordinate[below left = 1.5 cm of v1,label=left:$X_a$](i2);
\coordinate[above right = 1.5 cm of v1,label=right:$H$](i3);
\coordinate[below right = 1.5 cm of v1,label=right:$X_b$](i4);
\coordinate[below right = 1.5 cm of i1](v1);
\coordinate[above = 0.40 cm of v1,label=$\,Q$](QL);
\coordinate[below = 0.90 cm of v1,label=$\,U$](UL);
\coordinate[below=0.2cm of v1](v1aux);
\coordinate[right = 0.6 cm of v1aux,label=$\,t$](tR);
\coordinate[left = 0.55 cm of v1aux,label=$\,t$](tR2);
\draw[scalar](0.35,-0.35)--(i1);
\draw[scalar](i2)--(0.35,-1.7);
\draw[scalar](1.85,-0.35)--(i3);
\draw[scalar](i4)--(1.85,-1.7);
\draw[fermion](1.1,-1) circle (1);
\draw[fermionp](1.1,-1) circle (1);
\draw[fermionpp](1.1,-1) circle (1);
\draw[fermionppp](1.1,-1) circle (1);
\draw[fill=black] (1.85,-1.7) circle (.05cm);
\draw[fill=black] (0.35,-1.7) circle (.05cm);
\draw[fill=black] (0.35,-0.35) circle (.05cm);
\draw[fill=black] (1.85,-0.35) circle (.05cm);
\end{tikzpicture}
\end{gathered}
\end{equation*}
\caption{Some contributions to the terms $H^\dagger H$ and $H^\dagger H X_a^* X_b$ in the effective scalar potential.}
\label{fig:diagrams}
\end{figure}
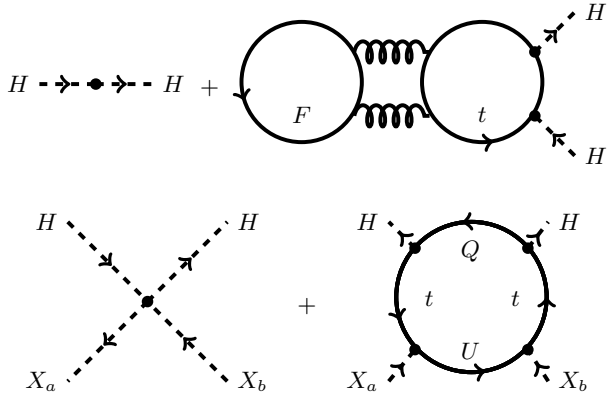
In a renomalizable theory after all divergencies are cancelled one can check if there are needed cancellation between different contributions to a physical quantify.
In order to avoid awkward fine tuning one can impose bounds on the masses and couplings of the new fields. This simple criteria is often called {\it finite naturalness}, 
see Ref.~\cite{Farina:2013mla} for a detailed discussion.

The Nelson-Barr mechanism is an attractive solution to the Strong CP problem. Without it, a very precise cancellation between the strong CP phase $\theta_{\rm QCD}$ and $\text{arg}\{ \text{det} ( {\cal M}_u {\cal M}_d) \}$ is needed to satisfy $\bar \theta_{\rm QCD} < 10^{-10}$. However, generically in Nelson-Barr models, new colored fermions, vector-like under the standard model, are required to implement the aforementioned mechanism. Moreover, because of the stable domain wall generated from the spontaneous breaking of a space-time symmetry as CP, $v_X$ must be large. Given the simultaneous presence of several scales in the theory, a similar cancellation problem to the one that Nelson-Barr models succeed to solve may arise in the effective scalar potential. Therefore, it is reasonable to explore for which coupling values the theory can remain free of such cancellations. 

The new vector-like fermions, as long as they carry standard model quantum numbers, will contribute at three loops to the Higgs mass, as shown in the upper panel of Fig.~\ref{fig:diagrams}. Particularly, for the new quarks~\cite{Farina:2013mla}, 
\begin{equation}
\delta m_H^2 \sim \frac{\alpha_s^2}{(4\pi)^4}M_F^2 \ln \left(\frac{M_F^2}{m_t^2}\right).
\end{equation}
This implies that
\begin{equation}
M_F \lesssim 500 \text{ TeV}\left( \frac{0.118}{\alpha_S}\right) \left( \frac{\delta m_H^2 / m_H^2}{100}\right)^{1/2}.
\label{eq:MFbound}
\end{equation}

Since inflating the domain walls away demands the VEV of $X$ to be large, one can assume that all couplings between the $X_i$ and the SM Higgs are suppressed, avoiding thus cancellations arising at tree-level in the scalar potential. For example, the coupling of the $H^\dagger H X_i X_j^*$ quartic interaction at tree level,
\begin{equation}\label{eq:treelevel}
\lambda <  10^{-14} \left(\frac{10^{10} \text{ GeV}}{v_X}\right)^2 \left(\frac{\delta m_H^2/m_H^2}{100}\right).
\end{equation}
This is consistent with having a large $\theta_X$, and an acceptably small contribution to $\bar \theta_{\rm QCD}$ from loops (see Eq.~\eqref{eq:1loop}).

On the other hand, such crossed-terms can be generated by loop effects. For instance, the diagram displayed in the bottom right panel of Fig.~\ref{fig:diagrams} contributes to the SM Higgs mass in the following way:
\begin{equation}\label{eq:42}
\delta m_H^2 \sim \frac{\lambda_u^2}{(4\pi)^2}v_X^2 \ln \left(\frac{M_U^2}{m_t^2}\right).
\end{equation}
Using the fact that $\lambda_u v_X \sim \tilde M_\textsc{U}$, 
Eq.~\eqref{eq:42} implies
\begin{equation} \label{eq:FiniteNat}
 M_{\textsc{U}} \lesssim 6 \text{ TeV} \left(\frac{\delta m_H^2 / m_H^2}{100}\right)^{1/2}.
\end{equation}

Eq.~\eqref{eq:FiniteNat} suggests that in this model vector-like fermions may be accessible to the LHC or future colliders. To achieve masses at the TeV scale with SCPV at a much larger scale (which is required by the domain walls) implies very small couplings.

There are other finite naturalness constraints which we will not discuss since there may be other physics  that makes tunings in the Higgs effective potential acceptable. 

\section{Summary}
In this article we have discussed simple gauge theories where the Nelson-Barr mechanism is realised. Using a gauge symmetry that is chiral under the new vector-like fermions improves the quality of the solution to the strong CP puzzle.

We studied models that correspond to ${\rm U}(1)_{R}$ when acting on the SM fermions. These models have CP violation in both the quark and lepton sectors. 

We derived explicit expressions for flavor violating interactions that apply in Nelson-Barr models which hold quite generally. These are more suppressed than one may naively expect, occurring at order $m^2/\tilde M_F^2$, where $m$ ($\tilde M_F$) is the magnitude of the standard model (new vector-like) fermion masses. This additional suppression ensures that they can be consistent with experimental constraints. 

If finite naturalness is taken seriously, it suggests that the new particle content is near the TeV scale. Another way that new particle content can be at the TeV scale is to have CP symmetry not restored at high temperature.

{\small {\textit{Acknowledgments}}: We thank Lisa Randall and Ryan Plestid for useful discussions. The work of C.M. and M.B.W. is supported by the U.S. Department of Energy, Office of Science, Office of High Energy Physics, under Award Number DE-SC0011632 and by the Walter Burke Institute for Theoretical Physics.}

\appendix

\section{Corrections to $\bar \theta_{\rm QCD}$}
Non-renormalizable operators and loops can contribute to $\bar \theta_{\rm QCD}$ by adding corrections to the tree-level mass matrix of the quarks. Then, the total mass matrix of the quarks, $\tilde {\cal M}_q$, can be decomposed into
\begin{equation}
\tilde {\cal M}_q = {\cal M}_q +  \delta {\cal M}_q,
\end{equation}
where ${\cal M}_q$ is the tree-level mass matrix of the quarks, see Eq.~\eqref{eq:Mf}, and $\delta {\cal M}_q$ parametrizes any sub-leading contribution. 

The determinant of a matrix close to the identity can be expanded as follows,
\begin{equation}
\begin{split}
\text{Det}(\tilde {\cal M}_q )&= \text{Det}({\cal M}_q) \, \text{Det}(\mathbb{I}+ {\cal M}_q^{-1} \delta {\cal M}_q)\\
&=\text{Det}({\cal M}_q) \, \left(1 + \text{Tr}\{ M_q^{-1}\delta {\cal M}_q\} \right),
\end{split}
\end{equation}
where we have employed the identity $\text{Det}(e^{A} )=  e^{\text{Tr}\{A\}}$.
Since the argument of a complex number is given by $\text{arg} \{z \} = -i (\ln z - \ln |z|)$, and using that 
\begin{equation}
\ln \text{Det} (\tilde {\cal M}_q) = \ln \text{Det} ({\cal M}_q) + \text{Tr}\{{\cal M}_q^{-1}\delta {\cal M}_q \},
\end{equation}
the following expression is derived,
\begin{eqnarray}
\text{arg}\{\text{Det} (\tilde {\cal M}_q )\} &= \text{arg} \{ \text{Det} ( {\cal M}_q) \}- i \, \text{Tr} \{ {\cal M}_q^{-1}\delta {\cal M}_q \} \nonumber \\
& \!\! + \, i \, \ln \left | 1 + \text{Tr}\{{\cal M}_q^{-1}\delta {\cal M}_q \}\right|. 
\end{eqnarray}
Using that $|z| = \sqrt{z^*z} $ and expanding the logarithm of the square root up to first order in $\delta {\cal M}_q$, the subleading corrections to the strong CP phase can be written as
\begin{equation}
\Delta \bar \theta_{\rm QCD} = \text{Im} \big \{\text{Tr}\{{\cal M}_q^{-1} \delta {\cal M}_q\} \big \} + {\cal O}(\delta M_q^2).
\label{eq:thetaQCDcorr}
\end{equation}

\bibliography{SCP}

\end{document}